\documentclass{soups} 
\usepackage{graphicx}
\usepackage{times}
\usepackage{times}
\usepackage{epsfig}
\usepackage{graphicx}
\usepackage{amsmath}
\usepackage{amssymb}
\usepackage{xfrac}
\usepackage{environ}
\usepackage{textcomp}
\usepackage{multirow}
\usepackage{hhline}
\usepackage{adjustbox}
\usepackage{wasysym}
\usepackage{balance}
\usepackage{color}
\usepackage{algorithm}
\usepackage[noend]{algpseudocode}

\begin{document}
%

\title{IllusionPIN: Shoulder-Surfing Resistant Authentication Using Hybrid Images\titlenote{This paper was published as \cite{papadopoulos2017illusionpin}}}
%
%
%
%
%

\numberofauthors{4} 
%
\author{
%
%
\alignauthor
Athanasios Papadopoulos\\
       \affaddr{New York University}\\
       \email{tpapadop@nyu.edu}
\alignauthor
Toan Nguyen\\
       \affaddr{New York University}\\
       \email{toan.v.nguyen@nyu.edu}
 \alignauthor Emre Durmus\\
      \affaddr{New York University}\\
       \email{mdurmus@nyu.edu}
\and \alignauthor Nasir Memon\\
       \affaddr{New York University}\\
       \email{memon@nyu.edu}
}


\maketitle
\begin{abstract}

We address the problem of shoulder-surfing attacks on authentication schemes by proposing IllusionPIN (IPIN), a PIN-based authentication method that operates on touchscreen devices. IPIN uses the technique of hybrid images to blend two keypads with different digit orderings in such a way, that the user who is close to the device is seeing  one keypad to enter her PIN, while the attacker who is looking at the device from a bigger distance is seeing only the other keypad. The user's keypad is shuffled in every authentication attempt since the attacker may memorize the spatial arrangement of the pressed digits.

To reason about the security of IllusionPIN, we developed an algorithm which is based on human visual perception and estimates the minimum distance from which an observer is unable to interpret the keypad of the user. We tested our estimations with 84 simulated shoulder-surfing attacks from 21 different people. None of the attacks was successful against our estimations. In addition, we estimated the minimum distance from which a camera is unable to capture the visual information from the keypad of the user. Based on our analysis, it seems practically almost impossible for a surveillance camera to capture the PIN of a smartphone user when IPIN is in use.

\end{abstract}

\section{Introduction}
%
%
%
User authentication is performed in various ways \cite{bonneau2012quest}. We focus on PIN authentication because of its simplicity and maturity. A Personal Identification Number (PIN) is a sequence of digits that confirms the identity of a person when it is successfully presented. PINs are simpler than alphanumeric  passwords as they solely consist of numerical characters (0-9) and have a short length that is usually either $4$ or $6$ digits. This makes PINs easy to remember and easy to reproduce, and as a consequence, PIN authentication is characterized by infrequent errors \cite{harbach2016anatomy}. So, simplicity is translated to usability. The maturity of PIN authentication is a result of its continuous usage for years in a wide range of everyday life applications, like mobile phones and banking systems. 

From the perspective of security, PIN authentication is susceptible to brute force or even guessing attacks \cite{bonneau1}. To balance this weakness, the number of allowed authentication attempts is usually constrained to a small number such as  $5$. However, a simple attack that is still very hard to mitigate is shoulder-surfing \cite{anderson1993cryptosystems}.

Shoulder-surfing refers to eavesdropping personal information, like an alphanumeric password or a PIN, through observation. A typical example is an adversary who is standing behind a person in the line for an ATM machine and is looking, or ``surfing", over the person's shoulder to obtain her PIN information. In this scenario, the attacker is observing a person while being in her vicinity. However, the attacker may observe someone remotely by using recorded material that was collected intentionally or even unintentionally. For example, unintentional recording of shoulder-surfing material could result from a surveillance camera that captured a person while entering her authentication credentials to unlock her phone in a store or at the workplace.

Authentication schemes which are not resilient to observation are vulnerable to shoulder-surfing. Any kind of visual information may be observed, including the blink of a button when it is pressed, or even the oily residue that the fingers leave on a touchscreen \cite{aviv2010smudge}. Shoulder-surfing is a big threat for PIN authentication in particular, because it is relatively easy for an observer to follow the PIN authentication process. PINs are short and require just a small numeric keypad instead of the usual alphanumeric keyboard. In addition, PIN authentication is often performed in crowded places, e.g., when someone is unlocking her mobile phone on the street or in the subway. Shoulder-surfing is facilitated in such scenarios since it is easier for an attacker to stand close to the user while escaping her attention.

The motivation behind this work relies on the hypothesis that PIN authentication will really meet the needs of its users when it will increase its shoulder-surfing resistance without a significant overhead in its usability. We contributed towards this claim in the following ways.

\begin{itemize}
  \item We designed IllusionPIN (IPIN) for touchscreen devices. The virtual keypad of IPIN is composed of two keypads with different digit orderings, blended in a single hybrid image \cite{oliva2006hybrid}. The user who is close to the screen is able to see and use one keypad, but a potential attacker who is looking at the screen from a bigger distance, is able to see only the other keypad. 
We analyze in detail the design of IllusionPIN in Section~\ref{sec:illusionPIN}.
  \item We developed an algorithm to estimate whether or not the user's keypad is visible to an observer at a given viewing position. We explain the estimation algorithm in Section~\ref{sec:visibility_algorithm}.
  \item We tested the estimated visibility of IllusionPIN through a user study of simulated shoulder-surfing attacks on smartphone devices. In total, we performed $84$ attacks with $21$ different people and none of the attacks was successful against our estimations. We provide the details of this user study in Section~\ref{sec:evaluation}.
  \item We estimated the minimum distance from which a camera is unable to capture the visual information from the user's keypad.
  The exact procedure is explained in Section~\ref{sec:safety_distance}. The results show that it is practically almost impossible for a surveillance camera to capture the PIN of a smartphone user when IllusionPIN is in use.
\end{itemize}

\section{Related Work}

We organize shoulder-surfing resistant authentication schemes according to $6$ design principles.
The \textbf{Obscurity} principle states that the visual information of interest has to be obscured. For example, ShieldPIN \cite{kim2010multi} requires the user to physically cover the keypad by cupping one hand while using the other hand to enter her PIN. It is obvious that such an approach demands physical effort and simultaneous usage of both hands that may be unwanted. An alternative solution that does not require any extra effort from the user is to make the content of the screen visible within a limited range of viewing angle. This can be achieved either with additional hardware, e.g. privacy filters, or with special hardware, e.g. automultiscopic displays \cite{chan2008top, matusik2008multiview}. In both cases, deployability may be an issue. However, there is a number of software solutions which create a similar effect \cite{harrison2011new, kim2012enabling}. Specifically, depending on the viewing angle, different visual elements appear on the screen and obscure the real content. These approaches exploit technical limitations of certain screens' technology, and as a result, they can't be generalized or expected to be applicable in the future as screen technology advances. In addition, a shoulder-surfer is not necessarily observing from a big angle, as he may be just standing behind the user.

The \textbf{Visual Complexity} principle states that it has to be difficult to receive the visual information of interest \cite{hayashi2008use}. For example, DAS \cite{jermyn1999design} is a simple graphical password that allows the user to create a free-form drawing on a touchscreen and to use it as her password. Decoy Strokes \cite{zakaria2011shoulder} is a shoulder-surfing resistant variation of DAS that draws strokes alongside the user's password to confuse a malicious observer. The problem with such schemes is that the user is exposed to the same distracting information and may end up confused as well, leading to slower authentication and more frequent input errors. Also, if the attacker is able to observe multiple times during the authentication process or to record it, he may be able to steal the credentials of the user.

The \textbf{Cognitive Complexity} principle states that it has to be difficult to process the acquired visual information \cite{tan2005spy}. For example, in one of the cognitive trapdoor games \cite{roth2004pin}, the user is required to enter her PIN in the following way. The digits on the provided keypad are separated into two sets based on their color; half of them are black, and half of them are white. The user selects the set that the current digit of her PIN belongs to, and then the digits are reassigned to the two color sets. This procedure is repeated until the scheme is able to uniquely determine the correct digit by intersecting the selected sets. Then, the user proceeds to the next digit of her PIN. For an observer, it is extremely difficult, if not impossible, to successfully perform sequential intersections of sets to extract the correct PIN. However, such schemes are usually complex for the users too, with all the inevitable consequences for usability. In addition, recorded material or even repeated observation may reveal the authentication credentials, since all the useful information is observable.

The \textbf{Alteration} principle states that the required input has to change in every authentication attempt \cite{pering2003photographic, yadav2015design}. For example, Deja Vu \cite{dhamija2000deja} presents to the user a number $n$ of images, and asks her to specify which of them belong to a predefined set of images, called the user's portfolio. In each authentication attempt, different images from the portfolio are assigned to the set of the $n$ images. As a consequence, an observer cannot learn the portfolio of the user in a limited number of observations. However, multiple observations may reveal the whole portfolio. In general, with such schemes is difficult for the user to get familiar with a standard input. For example, with Deja Vu the user has to identify different pictures in every authentication attempt. This requires additional cognitive effort and may affect the authentication time and the error rate.

The \textbf{One-to-Many} principle states that the same input has to correspond to more than one authentication credentials \cite{li2005association, wiedenbeck2006design, perkovic2010shoulder, de2010colorpin, roth2004pin, shi2009pin}. For example, SlotPIN \cite{kim2010multi} allows the user to enter a PIN by aligning four vertical reels of randomly ordered digits. The first reel is static and denotes the position of the first PIN digit. The other three reels get aligned by the user according to the PIN. In the end, ten PINs are formed, and a shoulder-surfer is unable to know which is the correct one. In addition, it is difficult to memorize all of them. However, the attacker could replicate the same input without the need to know the correct credentials. That's why schemes designed under this principle randomize the input interface periodically. In the case of SlotPIN, the digits on the reels are randomized in every authentication attempt. The problem with such schemes is that they exhibit high complexity in order to break the one-to-one correspondence between inputs and authentication credentials. This results in high cognitive effort on the user side and renders such schemes unacceptable for frequent usage, e.g. for unlocking mobile phones. Also, multiple observations may reveal the correct credentials.

The \textbf{Non-Visual Information} principle states that at least part of the information of interest has to be transmitted through channels that are not observable. This way, an observer is always missing a piece of information and shoulder-surfing is mitigated even in cases that multiple observations or recordings are possible. However, the performance of such schemes in usability and deployability varies. For example, schemes which use audio and haptic information \cite{bianchi2011phone, sasamoto2008undercover} suffer from high authentication time and their requirements for additional hardware heavily affect their deployability. However, the emergence of touchscreen devices which are pressure-sensitive may favor schemes which use this kind of haptic information \cite{malek2006novel, sae2012biometric, nguyen2014finger, sae2014multitouch, sae2014online, nguyen2017draw}, if they are combined with satisfying usability. Gaze-based authentication \cite{kumar2007reducing} corresponds to high authentication time, and authentication based on brainwaves \cite{chuang2013think} or bone conduction \cite{schneegass2016skullconduct} requires special equipment. Fingerprint authentication \cite{ratha2001enhancing} is a scheme that gains popularity nowadays by offering excellent usability. A usual problem with biometrics like fingerprints is that they cannot be revoked. In addition, biometrics can be used to uniquely identify a person and they raise privacy concerns. However, the concept of cancelable biometrics \cite{patel2015cancelable} could alleviate both problems.

\section{Threat Model}
A threat model guides the design of a security scheme. Our threat model consists of shoulder-surfing attack scenarios against smartphone authentication. Different scenarios involve attacks with different difficulty to mitigate. The most difficult scenario that we consider corresponds to situations similar to a crowded train where we assume that a shoulder-surfer may be as close as $25$ inches to the user. The least dangerous scenario we consider corresponds to workplace conditions, where a shoulder-surfer has the ability to repeatedly observe the user but from a minimum distance of $60$ inches (e.g. the shoulder-surfer works to the cubicle next to the user). A scenario with intermediate difficulty corresponds to a non-crowded public place, where a shoulder-surfer may approach the user in a radius of $35$ or $45$ inches. As we can see, in each scenario we have to protect the user for viewing distances which are equal or bigger than a particular distance. We call such a distance ``safety distance", and we denote it with $d_s$. It is also important to consider that when the shoulder-surfer stands behind the user, it is difficult to have visual contact with the phone screen even if he is considerably taller than the user. Based on that, we assume that the shoulder-surfer is standing next to the user at an angle that is at least $30$ degrees. We would like to note that the aforementioned safety distance values that ranged from $25$ to $60$ inches and corresponded to different scenarios, were determined empirically. Similarly, the value of $30$ degrees is empirical too.

We additionally consider the scenario that the attacker has the ability to record the user during the authentication process. It is difficult for an attacker to record a user from a close distance while escaping her attention. Also, it is difficult to capture the user during the short time of the authentication process that usually happens unexpectedly. Consequently, we will focus on the scenario that the recorded material is collected with surveillance cameras. In such a case, the distance of recording is assumed to be at least $100$ inches.

\section{Perception of an Image by a Human Observer}
To explain how we designed IllusionPIN we first need to provide some background information about the perception of grayscale images. Based on the 2D Fourier transform which describes an image as a superimposition of sine wave gratings, we first examine the perception of a single sine wave grating and then we extend to arbitrary images.

\subsection{Perception of Sine Wave Gratings}
\begin{figure}[!t]
\centering
\includegraphics[width=3.5in]{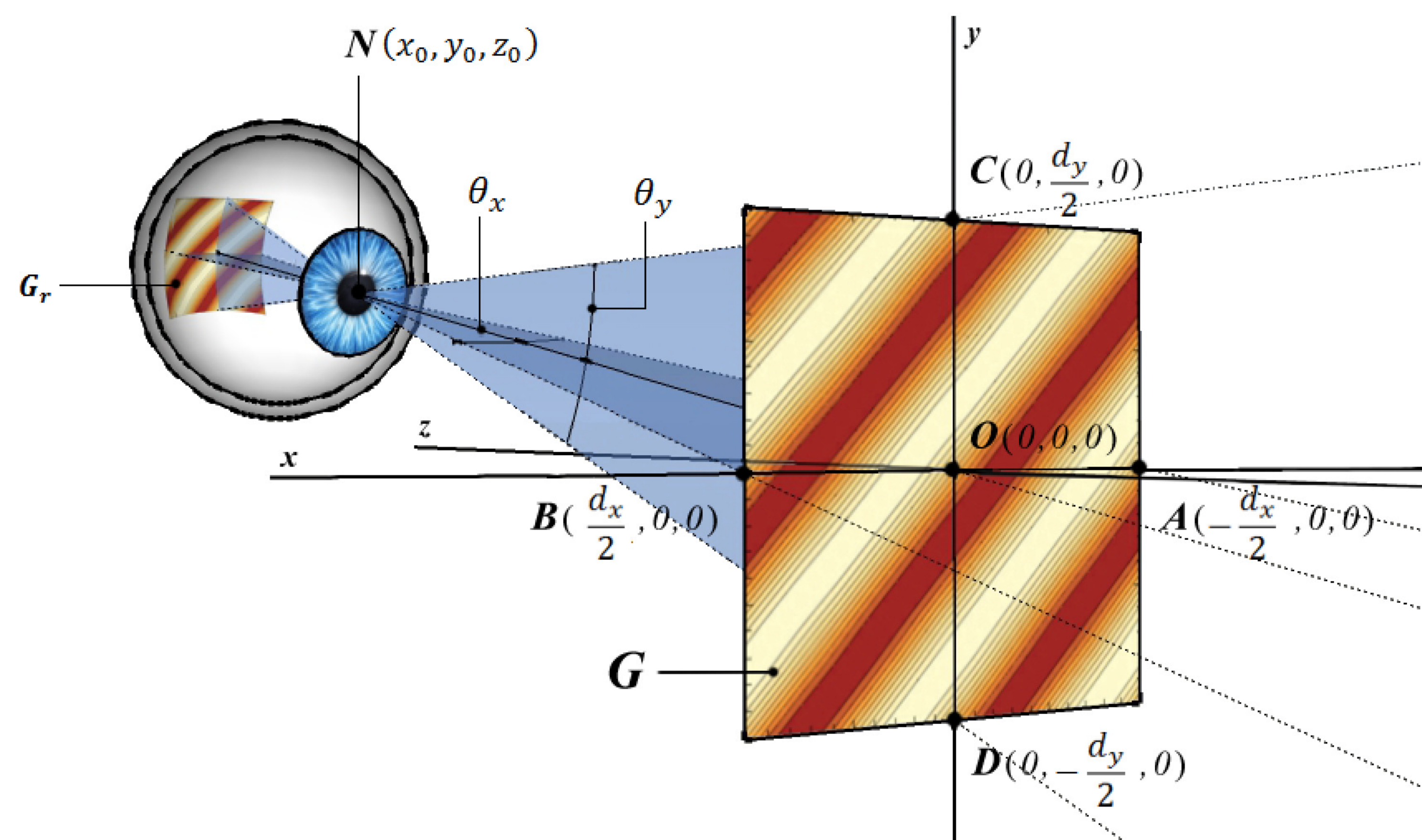}
\caption{The image formation process on the retina according to the model of the pinhole camera. $G_r$ is the projection of image $G$ on the retina when is viewed from position $N$. Spatial positions are specified according to the coordinate system that is placed at the center of $G$.}
\label{fig:fig_5}
\end{figure}
We study how a human observer perceives an image of a sine wave grating by studying how this is done by a single eyeball. In figure \ref{fig:fig_5}, we show how we model the behavior of an eyeball. We assume that it behaves as an ideal pinhole camera which directs light rays from the surface of an image $G$ through a pinhole and onto the retina. On the retina, photoreceptors get activated and form a different 2D image $G_r$ which corresponds to the visual information that the eye receives. This is a simplification of the real process, but offers an acceptable approximation, while being mathematically convenient. Under this model, $G_r$ is shaped through the perspective projection of $G$ on the retina, and consequently, $G_r$ is a scaled version of $G$. This means that $G$ and $G_r$ are both sine wave gratings, but  with different parameters, meaning different spatial frequency, contrast and phase. So, we can think of the image formation process on the retina as a projection that modifies the parameters of $G$. This leads us to make a distinction between the actual and the perceived parameters of $G$, which correspond to its parameters before and after its projection on the retina respectively. The perceived parameters of $G$ are the parameters of $G_r$. Based on these remarks, given the actual parameters of a grating $G$ and the position $N$ of an observer, we want to derive the perceived parameters of $G$.

We start by considering the relation between the actual and  perceived contrast. Perceived contrast depends on the amount of light rays that reaches the retina from each part of an image, and consequently, we need to model the illumination of the 3D scene. In the context of our work, illumination results from the screen of the device, from external light sources, e.g. a light bulb, and from light reflected off surfaces in the surrounding space. We make two assumptions regarding illumination. The first is that illumination is uniform. This means that each part of the image emits the same amount of light. This is a fair assumption since we consider that the dominant light source is the screen of the device. The second assumption is that the amount of light rays that reaches the retina is adequate to perceive the visual information of the image. This assumption is based on the ability of smartphone devices to adjust the brightness of their screens to the illumination level of the surrounding space. Following these assumptions, we lift the need to make a distinction between the actual and the perceived contrast of a grating, and from now on we will consider them as being {equal}.

Next, we consider perceived phase. As we have seen, $G_r$ is a scaled version of $G$. Scaling preserves the number of cycles of a grating as well as its phase, and consequently, we will consider perceived phase to be equal to the actual phase.

Now we move to the calculation of the perceived spatial frequency. In figure \ref{fig:fig_5}, we can see that $G$ subtends visual angle $(\theta_x, \theta_y)$. So, if $\vec{f} = (\sfrac{n_x}{d_x}, \sfrac{n_y}{d_y})$ is the actual spatial frequency of G, its perceived spatial frequency will be $\vec{f_p} = (\sfrac{n_x}{\theta_x}, \sfrac{n_y}{\theta_y})$, where $n_x$, $n_y$ is the number of cycles in the horizontal and vertical directions, and $(d_x, d_y)$ is the size of the image measured in units of length. All we need to do is to calculate $(\theta_x, \theta_y)$. To this end, we consider the general setting depicted in figure \ref{fig:fig_5}. The coordinate system for specifying spatial positions is placed at the center of the image where we assume that the observer is focused. The position of the observer is $N = (x_0, y_0, z_0)$, where $z_0 > 0$, meaning that the observer is in front of the image. For $\theta_x$ we have:
\begin{equation} \label{eq:1}
\theta_x = \cos^{-1} \Bigg( \frac{\overrightarrow{NA} \cdot \overrightarrow{NB}}{|\overrightarrow{NA}| \cdot |\overrightarrow{NB}|} \Bigg)
\end{equation}
For $\overrightarrow{NA}$ we have:
\begin{equation} \label{eq:2}
\begin{split}
\overrightarrow{NA} &= \overrightarrow{NO} + \overrightarrow{OA}\\
&= -(x_0, y_0, z_0) + (-\sfrac{d_x}{2}, 0, 0)\\
&= (-x_0-\sfrac{d_x}{2}, -y_0, -z_0)
\end{split}
\end{equation}
Similarly for $\overrightarrow{NB}$ we have:
\begin{equation} \label{eq:3}
\overrightarrow{NB} = (-x_0+\sfrac{d_x}{2}, -y_0, -z_0)
\end{equation}
Based on equations \ref{eq:2} and \ref{eq:3}, from equation \ref{eq:1} we get:
\begin{equation} \label{eq:7}
\resizebox{0.9\hsize}{!}{
$\theta_x = \cos^{-1} \Bigg( \frac{x_0^2 - \sfrac{1}{4} \cdot d_x^2 + y_0^2 + z_0^2}{\sqrt{(x_0+\sfrac{d_x}{2})^2 + y_0^2 + z_0^2} \cdot \sqrt{(x_0-\sfrac{d_x}{2})^2 + y_0^2 + z_0^2}} \Bigg)$
}
\end{equation}
Similarly, for $\theta_y$ we get:
\begin{equation} \label{eq:8}
\resizebox{0.9\hsize}{!}{
$\theta_y = \cos^{-1} \Bigg( \frac{y_0^2 - \sfrac{1}{4} \cdot d_y^2 + x_0^2 + z_0^2}{\sqrt{(y_0+\sfrac{d_y}{2})^2 + x_0^2 + z_0^2} \cdot \sqrt{(y_0-\sfrac{d_y}{2})^2 + x_0^2 + z_0^2}} \Bigg)$
}
\end{equation}
From equations \ref{eq:7} and \ref{eq:8} we can calculate $\theta_x$ and $\theta_y$.

If we transform equations \ref{eq:7} and \ref{eq:8} in spherical coordinates $(r, \theta, \phi)$, where $r \in (0, +\infty)$ is the viewing distance, $\theta \in (0, \pi)$ is the polar angle, and $\phi \in (-\sfrac{\pi}{2}, +\sfrac{\pi}{2})$ is the azimuth angle, for $N = (r_0, \theta_0, \phi_0)$ we get:
\begin{equation} \label{eq:12}
\resizebox{0.88\hsize}{!}{
$\theta_x = \cos^{-1} \Bigg( \frac{r_0^2 - \sfrac{1}{4} \cdot d_x^2}{\sqrt{(r_0^2 + \sfrac{1}{4} \cdot d_x^2)^2 - (\sfrac{1}{2} \cdot r_0 \sin{\theta_0} \sin{\phi_0} d_x)^2}} \Bigg)$
}
\end{equation}
\begin{equation} \label{eq:13}
\resizebox{0.88\hsize}{!}{
$\theta_y = \cos^{-1} \Bigg( \frac{r_0^2 - \sfrac{1}{4} \cdot d_y^2}{\sqrt{(r_0^2 + \sfrac{1}{4} \cdot d_y^2)^2 - (\sfrac{1}{2} \cdot r_0 \cos{\theta_0} d_y)^2}} \Bigg)$
}
\end{equation}
We can use these equations to examine how viewing distance and viewing angle affect visual perception. If we vary the value of viewing distance $r_0$, $\theta_x$ and $\theta_y$ can take all their possible values. When $r_0 = 0$ we get $\theta_x = \theta_y = \pi$. When $r_0$ approaches infinity, the factor $r_0^2$ dominates equations \ref{eq:12} and \ref{eq:13}, and we get $\theta_x \approx \theta_y \approx \cos^{-1} \frac{r_0^2}{\sqrt{r_0^4}} = 0$. In contrast, when we change the values of $\theta_0$ or $\phi_0$, we just affect the denominators in equations \ref{eq:12} and \ref{eq:13} and the value of visual angle still heavily depends on the viewing distance.
This shows that viewing distance is the main factor that affects visual perception. For this reason, in the following sections, when we need to demonstrate changes in perception we mainly consider variations in the viewing distance of the observer. However, to accurately estimate the way an image is perceived we have to consider the exact viewing position of the observer.

\subsection{Contrast Sensitivity Function}
\begin{figure}[!t]
\centering
\includegraphics[width=1.7in]{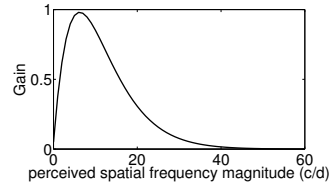}
\caption{The Contrast Sensitive Function model as proposed by Mannos et al. \protect\citeA{mannoseffects}.}
\label{fig:fig_9}
\end{figure}
The human visual system (HVS) demonstrates variability in its sensitivity to the perceived contrast of {gratings with different perceived spatial frequencies}. This variability is modeled through the contrast sensitivity function (CSF), which can be seen as a band-pass filter that the HVS applies to gratings according to their perceived spatial frequencies. Figure \ref{fig:fig_9} provides the model of the CSF proposed by Mannos et al. \cite{mannoseffects}. CSF favors gratings with perceived spatial frequency magnitudes at a particular range around $10 \; \sfrac{c}{d}$. For example, a grating with perceived spatial frequency magnitude equal to $30 \; \sfrac{c}{d}$ must have bigger perceived contrast than a grating with perceived spatial frequency magnitude $10 \; \sfrac{c}{d}$ to be perceived with the same clarity. Gratings with perceived spatial frequency magnitude beyond the limit of the visual acuity cannot be perceived, and that's why CSF cuts off completely every perceived spatial frequency with magnitude over $60 \; \sfrac{c}{d}$.

\subsection{Perception of Superimposed Sine Wave Gratings}

\begin{figure}[!t]
\centering
\includegraphics[width=3.5in]{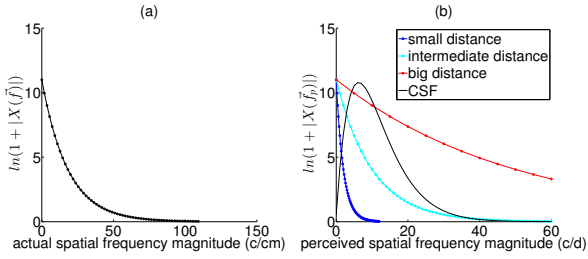}
\caption{(a) An example of how the actual spectrum of an image looks like. (b) The corresponding perceived spectrums when the same image is directly viewed from $3$ different viewing distances.}
\label{fig:fig_13}
\end{figure}

According to the 2D Fourier transform, an image {can be seen as} a superimposition of sine wave gratings with unique actual spatial frequencies. The contribution of each grating is quantified by the magnitude of its Fourier coefficient $|X(\vec{f})|$, which is proportional to the grating's amplitude by a constant factor. Given an example image, in figure \ref{fig:fig_13} (a) we visualize its actual 1D spectrum\footnote{The diagrams in figure \ref{fig:fig_13} do not originate from an existing image. The reason is that we wanted to make the demonstrations simpler (smoother curves) in order to be more understandable.}. The horizontal axis represents the magnitudes of actual spatial frequencies to create a simple 2D graph instead of a 3D one. {Unless stated otherwise, we will refer to 1D spectrums simply as spectrums.} The vertical axis represents the contribution of gratings through the quantities $\ln{(1+|X(\vec{f})|)}$. When the actual spatial frequencies $\vec{f_i}, i = 1,2,...n$ from $n$ different gratings have the same magnitude $|\vec{f_1}|$, the corresponding $\ln{(1+|X(\vec{f_i})|)}$ values are added together and the result is considered the total contribution of the $n$ gratings. Since we are dealing with digital images, we are using the Discrete Fourier Transform (DFT) and the circles on the curve of figure \ref{fig:fig_13} (a) correspond to the existing discrete magnitude values of actual spatial frequencies.

To understand how an image is perceived from a specific viewing position $N = (r_0, \theta_0, \phi_0)$, we use equations \ref{eq:12} and \ref{eq:13} to express the actual spectrum in perceived spatial frequencies. We call such a diagram the perceived spectrum. To aid our demonstrations, we consider the special case that the observer is looking directly at the image ($\theta_0 = \sfrac{\pi}{2}, \phi_0 = 0$) and additionally holds that $d_x = d_y = d$. In such a case, from equations \ref{eq:12} and \ref{eq:13} we get $\theta_x = \theta_y = \theta$. As a consequence, it holds that $|\vec{f_p}| = \sfrac{d}{\theta} \cdot |\vec{f}|$ and the perceived spectrum has exactly the same form as the actual spectrum. In figure \ref{fig:fig_13} (b), we depict the perceived spectrums of the image with the actual spectrum of figure \ref{fig:fig_13} (a) when it is viewed from $3$ different distances $r_0$. The blue curve corresponds to the smallest viewing distance. As the viewing distance increases, the visual angle gets smaller and the factor $\sfrac{d}{\theta}$ gets bigger. As a consequence, the perceived spectrums are stretched and take the form of the cyan and the red curve. Since the contrast of a grating can be expressed through its amplitude, we can use the CSF to filter the perceived spectrums in order to understand how they are perceived. As we can see, the band-pass nature of the CSF favors the cyan curve. In particular, the CSF assigns small gain values to gratings with a big contribution in both the blue and the red curve. As a consequence, part of the visual information that the image is carrying is either perceived with less clarity or it is not perceived at all. For example, in the case of the red curve that the viewing distance has its bigger value, many gratings with high perceived spatial frequency magnitudes are completely cut-off. That's why when an image is viewed from a big distance, it is perceived as being blurred.

\section{IllusionPIN (IPIN)}
\label{sec:illusionPIN}

\subsection{Method}
\begin{figure}[!t]
\centering
\includegraphics[width=3.5in]{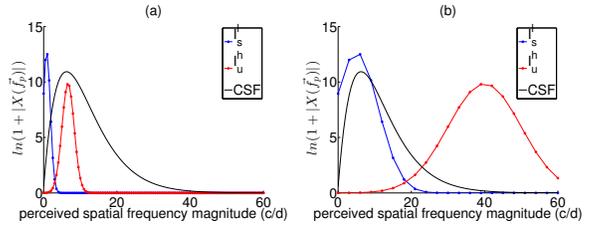}
\caption{(a) An example of how the perceived spectrums of two square images $I_u^h$ and $I_s^l$ look like when they are directly viewed from a small distance. $I_u^h$ is the result of high-pass filtering and $I_s^l$ of low-pass filtering. (b) The perceived spectrums of the same images when they are viewed from a bigger distance.}
\label{fig:fig_12}
\end{figure}
IllusionPIN is a PIN-based authentication scheme for touchscreen devices which offers shoulder-surfing resistance. The design of IllusionPIN is based on the simple observation that the user is always viewing the screen of her device from a smaller distance than a shoulder-surfer. Based on this, the core idea of IllusionPIN is to make the keypad on the touchscreen to be interpreted with a different digit ordering when the viewing distance is adequately large. This way, when the shoulder-surfer is standing far enough, he is viewing the keypad as being different from the one that the user is utilizing for her authentication, and consequently he is unable to extract the user's PIN. Also, the keypad is shuffled in every authentication attempt (or every digit entry) to avoid disclosing the spatial distribution of the pressed digits.
We create the keypad of IllusionPIN with the method of hybrid images \cite{oliva2006hybrid, oliva2013art} and we call it a hybrid keypad.

\begin{figure}[!t]
\centering
\includegraphics[width=3.5in]{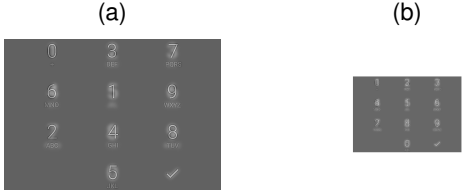}
\caption{(a) A hybrid keypad (b) The simulated perception of the same hybrid keypad when it is directly viewed from distance which is $2$ times bigger.}
\label{fig:fig_3}
\end{figure}

A hybrid keypad $I$ is created by blending appropriately two normal keypads denoted with $I_u$ and $I_s$. Our goal is $I$ to be interpreted as being $I_u$ when it is viewed from close up, and to be interpreted as being $I_s$ when it is viewed from far away. That's why {we call $I_u$ the ``user's keypad" and $I_s$ the ``shoulder-surfer's keypad"}. To create $I$, we process $I_u$ with a high-pass filter and $I_s$ with a low-pass filter. The filtering results in two new images, $I_u^h$ and $I_s^l$, and we simply set $I = I_u^h + I_s^l$. So, a hybrid keypad is composed by a high spatial frequency component $I_u^h$ and a low spatial frequency component $I_s^l$. To understand how the interpretation of $I$ is changing, we consider that we directly view an example hybrid keypad from different distances. If the viewing distance is adequately small, the visual angle is such that the perceived spectrums of $I_u^h$ and $I_s^l$ occupy low perceived spatial frequency magnitudes as depicted in figure \ref{fig:fig_12} (a). As we can see, the gain distribution of the CSF favors the perceived spectrum of $I_u^h$, and as a result $I$ is interpreted as being $I_u$. In figure \ref{fig:fig_12} (b) we depict the same perceived spectrums for a bigger viewing distance. As we can see, the perceived spectrums are stretched to higher perceived spatial frequency magnitudes and the CSF favors $I_s^l$. As a consequence, $I_s$ dominates the perception of $I$. From intermediate viewing distances, both $I_u^h$ and $I_s^l$ are visible to a considerable extent and is not certain how $I$ is interpreted. In figure \ref{fig:fig_3}, we provide an example hybrid image. In figure \ref{fig:fig_3} (b) we downscale the image of figure \ref{fig:fig_3} (a) to simulate how it is perceived when it is directly viewed from a $2$-times bigger distance. From reading distance, we expect the digit ordering of the hybrid keypad in figure \ref{fig:fig_3} (a) and (b) to be perceived as being different.

\subsection{Parameters}
Given the images $I_u$ and $I_s$, the parameters that specify the hybrid keypad $I$ are the parameters of the low-pass and the high-pass filters which are used to create $I_u^h$ and $I_s^l$.
For the low-pass filtering, we use a 2D Gaussian filter $G_l$, and for the high-pass filtering we use a filter $G_h = 1 - G_l^h$, where $G_l^h$ is also a 2D Gaussian filter. In the spatial domain, both $G_l$ and $G_l^h$ have zero mean values and diagonal covariance matrices with equal standard deviations. This way, they affect an image by an equal number of pixels in each dimension. So, if $\sigma^{ls}_x$ and $\sigma^{ls}_y$ are the standard deviations of $G_l$ in the horizontal and vertical dimensions, it will hold $\sigma^{ls}_x = \sigma^{ls}_y$. If $\sigma^{lf}_x$ and $\sigma^{lf}_y$ are the corresponding standard deviations in the frequency domain, it will hold $\sfrac{N_x^f}{2 \pi \sigma^{lf}_x} =  \sfrac{N_y^f}{2 \pi \sigma^{lf}_y} \Rightarrow \sigma^{lf}_y = \sfrac{N_y^f}{N_x^f} \cdot \sigma^{lf}_x$, where $(N_x^f, N_y^f)$ is the size of the filter in samples. This shows that we can completely define $G_l$ by specifying either $\sigma^{ls}_y$ or $\sigma^{lf}_y$. For convenience, we prefer to work in the frequency domain and consequently, we consider $\sigma^{lf}_y$ as the single parameter of $G_l$. Similarly, we define $G_h$ through the standard deviations of $G_l^h$, for which hold $\sigma^{hf}_y= \sfrac{N_y^f}{N_x^f} \cdot \sigma^{hf}_x$ in the frequency domain. Consequently, we consider $\sigma^{hf}_y$ as the single parameter of $G_h$.

We have to make two additional remarks. The first is that we want a filtered image to maintain its original size in pixels, $(N_x, N_y)$, without the creation of additional frequency components in the stop-band of the filters. To this end, we perform the filtering in the frequency domain by multiplying the $(N_x, N_y)$-point DFT of the image with the corresponding filter. Because of this, the ratio $\sfrac{N_y^f}{N_x^f}$ is equal to the ratio of the image dimensions $\sfrac{N_y}{N_x}$, which usually is $\sfrac{16}{9}$. The second remark is that we measure $\sigma^{lf}_y$ and $\sigma^{hf}_y$ in number of cycles per image ($\sfrac{c}{im}$). This way, the visual effect of filtering is invariant to the size of the image.

\subsection{Tuning}
The values of $\sigma^{lf}_y$ and $\sigma^{hf}_y$ create a trade-off between security and usability. When the value of $\sigma^{lf}_y$ is increased, $I_s^l$ becomes more visible since its perceived spectrum extends to higher spatial frequencies. This means that the user gets more distracted during her authentication and she may even need to bring the touchscreen closer to her eyes to clearly see the user's keypad. So, usability is negatively affected. However, a shoulder-surfer needs to reduce his viewing distance too in order to see the user's keypad with the same clarity, and consequently security is increased. When $\sigma^{lf}_y$ is decreased, the opposite effects are caused and usability is increased while security is decreased. The same trade-off is observed when the value of $\sigma^{hf}_y$ changes, since the clarity of $I_u^h$ is affected. In particular, when $\sigma^{hf}_y$ is decreased, usability is increased and security is decreased, while the opposite behavior is observed when $\sigma^{hf}_y$ is increased.

To resolve the trade-off between security and usability, we first set the value of $\sigma^{lf}_y$ in such a way that every possible level of security and usability is still possible depending on the value of $\sigma^{hf}_y$. Then, based on our security requirements, we set the minimum $\sigma^{hf}_y$ that satisfies them. This way, usability is maximized under the constraint that our security needs are met. To set the value of $\sigma^{lf}_y$, we consider that if $\sigma^{lf}_y$ gets very small, $I_s^l$ will be that blurred that the original digits from $I_s$ won't be recognizable, no matter the viewing distance. On the other hand, we have to consider that a user is usually holding her device at a particular distance. If $\sigma^{lf}_y$ gets very big, the user's keypad won't be able to dominate the perception from that specific viewing distance. So, we set $\sigma^{lf}_y$ to be close to the minimum value that allows the digits on $I_s^l$ to be interpreted. We experimented with $Nexus \; 6$ and $iPhone \; 6$ smartphones which have representative keypads at their lock screens while they differ in size, resolution, and visual content. We concluded that a suitable value for $\sigma^{lf}_y$ is $35 \; \sfrac{c}{im}$. However, if the digits in another keypad are considerably different, e.g. much thicker, we may need to adjust the value of $\sigma^{lf}_y$ to make them equally recognizable. 

To specify the value of $\sigma^{hf}_y$ we have to consider the given security requirements, which correspond to a safety distance value. In Section~\ref{sec:visibility_algorithm} we explain how we estimate the minimum value of $\sigma^{hf}_y$ that respects a particular safety distance.

\subsection{Discussion}
For the design of IllusionPIN we follow the principle of obscurity, since the shoulder-surfer's keypad $I_s$ obscures the user's keypad $I_u$. We could use any image in the place of $I_s$, but we decided to always use {the image of keypad} because this way $I_u^h$ and $I_s^l$ are visually aligned. This means that the digits in $I_u^h$ and $I_s^l$ overlap, and the less dominant keypad is perceived as noise, providing more clear interpretations of the hybrid keypad. 

We also apply the principle of alteration by shuffling the user's keypad in every authentication attempt, or after every digit entry. Otherwise, it would be enough for a shoulder-surfer to memorize just the spatial arrangement of the pressed digits. However, the shoulder-surfer's keypad always remains the same because this way we expect the user to become gradually better on ignoring it, resulting in faster authentication with fewer errors. Out of all the possible digit orderings that $I_s$ may have, we choose the regular digit ordering, as in figure \ref{fig:fig_3}. The reason is that this is the ordering that we expect an attacker to be the most familiar with, and as a consequence, to have the tendency to recognize.

In our threat model, we have considered 4 shoulder-surfing scenarios with safety distance values which {are} equal to $25$, $35$, $45$ and $60$ inches. For each of these values we estimate the minimum $\sigma^{hf}_y$ that keeps the user protected. This way we create $4$ hybrid keypads for which hold that when security is increased, usability is decreased. These keypads are offered as predefined options to the user to pick the one that fits better to her needs.

\begin{algorithm}[!t]
\caption{Visibility Algorithm}\label{euclid}
\textbf{Require:} hybrid keypad $I$, shoulder-surfer's keypad $I_s$, $DAF$ filter, viewing position $N$, visibility index threshold value $v_{th}$
\begin{algorithmic}[1]
\State $I_{sp} \gets \text{calculate $2D$ perceived spectrum of $I$ given $N$}$
\State $I_{s,sp} \gets \text{calculate $2D$ perceived spectrum of $I_s$ given $N$}$
\State $I^{DAF}_{sp} \gets \text{apply DAF filter to $I_{sp}$}$
\State $I^{DAF}_{s,sp} \gets \text{apply DAF filter to $I_{s,sp}$}$
\State $I^{DAF} \gets \text{transform $I^{DAF}_{sp}$ to spatial domain}$
\State $I^{DAF}_{s} \gets \text{transform $I^{DAF}_{s,sp}$ to spatial domain}$
\State $Buttons(I^{DAF}) \gets \text{segment the buttons of $I^{DAF}$}$
\State $Buttons(I^{DAF}_{s}) \gets \text{segment the buttons of $I^{DAF}_{s}$}$
\State $v \gets \text{mean}(\text{MSSIM}(Buttons(I^{DAF}), Buttons(I^{DAF}_{s})))$
\If {$ v \geq v_{th}$}
\State $\text{\textbf{return}} \; \text{No}$
\Else
\State $\text{\textbf{return}} \; \text{Yes}$
\EndIf
\end{algorithmic}
\end{algorithm}

\section{Visibility Algorithm}
\label{sec:visibility_algorithm}
The visibility algorithm receives as inputs a hybrid keypad $I$ and a viewing position $N$ in the 3D space. It returns a {binary} prediction on whether the user's keypad of $I$ is visible to an observer who is in position $N$. We use this prediction either to estimate the minimum safety distance that corresponds to a given hybrid keypad, or to create a hybrid keypad that respects a given safety distance. {Algorithm $1$ provides the pseudocode of the visibility algorithm.}

\subsection{Algorithm}
\subsubsection{Distance-As-Filtering}
\label{subsubsec:distance}
In the first step of the visibility algorithm, we simulate the way $I$ is perceived from the viewing position $N$ by using the distance-as-filtering hypothesis proposed by Loftus et al. \cite{loftus2005easier}. The distance-as-filtering hypothesis states that we can simulate the way an image is perceived from a particular viewing distance by filtering the image with an appropriate low-pass filter. The intuition behind this method is based on the effect that the CSF has in the visual perception of an image. However, as explained by Loftus et al. \cite{loftus2005easier}, the perception of an image from a particular distance and the perception of the corresponding filtered image are not identical, but they are equivalent with respect to  performance on some task. Loftus et al. experimentally verified the distance-as-filtering hypothesis for face recognition tasks by designing a low-pass filter of a particular form. Our task is the recognition of digits on hybrid keypads, which differs from face recognition. However, both tasks require the perception of almost equally fine visual details, and consequently, we expect the low-pass filter designed by Loftus et al. to be applicable in our task too. We should also note that in the experiments conducted by Loftus et al., every observer was looking directly at an image and his viewing position was completely defined by his viewing distance. We  use the same filter to simulate the perception of an observer who is at a random viewing position, {by making} the simplifying assumption that the perception of an image depends on the visual angle that it subtends, no matter where the observer stands.

The low-pass filter proposed by Loftus et al. \cite{loftus2005easier} has constant gain equal to $1$ until the perceived spatial frequency magnitude $|\vec{f_0}|$, and then drops until it reaches the value $0$ at the perceived spatial frequency magnitude $|\vec{f_1}|$. We call it DAF filter and it is mathematically defined in the following way:
\begin{equation} \label{eq:14}
DAF(\vec{f_p}) =
\left\{
	\begin{array}{ll}
		1  & \mbox{if} |\vec{f_p}| < |\vec{f_0}| \\
		1 - \Big( \frac{\log{(\sfrac{|\vec{f_p}|}{|\vec{f_0}|})}}{\log{(r)}} \Big)^2 & \mbox{if} |\vec{f_0}| \leq |\vec{f_p}| \leq |\vec{f_1}| \\
		0  & \mbox{if} |\vec{f_p}| > |\vec{f_1}|
	\end{array}
\right.
\end{equation}
where $r$ is a positive constant for which holds $r > 1$, and $|\vec{f_0}| = \sfrac{|\vec{f_1}|}{r}$. So, the parameters of the filter are the values of $r$ and $|\vec{f_1}|$. Loftus et al. conducted $4$ different face recognition experiments to specify the values of the parameters, and concluded that $r = 3$, while $|\vec{f_1}|$ may be equal to $25$, $31$ or $42 \; \sfrac{c}{d}$, depending on the task at hand. In Section~\ref{subsubsec:DAF}, we explain how we specified $|\vec{f_1}|$ for the purposes of our work. 

\subsubsection{Visibility Index}
\begin{figure}[!t]
\centering
\includegraphics[width=3.5in]{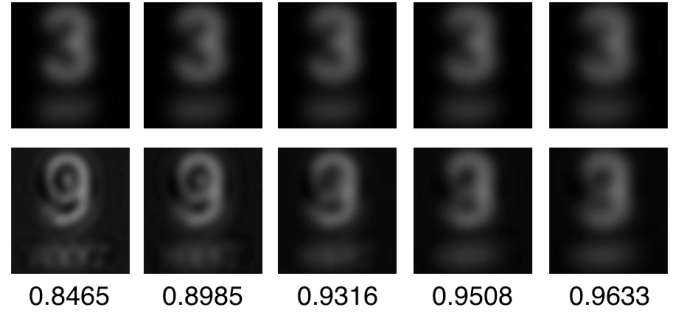}
\caption{{We consider an example} hybrid keypad $I = I_s^l + I_u^h$. In the first row,  the third button of $I_s^{l, DAF}$ is depicted when $I_s^l$ is directly viewed from different distances. In the second row, the corresponding button of $I^{DAF}$ is depicted when $I$ is directly viewed from the same distances as $I_s^l$. In the third row, the value of the visibility index for each viewing distance is provided.}
\label{fig:fig_22}
\end{figure}
In the second step of our algorithm, we compute the visibility index which quantifies how visible the user's keypad of $I$ from the viewing position $N$ is. We remind that $I = I_s^l + I_u^h$. To compute the visibility index, we apply the DAF filter both to $I$ and to $I_s^l$, and we create the images $I^{DAF}$ and $I_s^{l, DAF}$, respectively. This way we simulate how $I$ and $I_s^l$ are perceived when they are viewed from position $N$. Then, we separate in equal rectangular regions the buttons from $I^{DAF}$ and $I_s^{l, DAF}$, and we compute the similarity of the corresponding buttons with the mean structural similarity index (MSSIM) \cite{wang2004image}. The visibility index is the mean value of the $10$ MSSIM index values from the pairs of corresponding buttons.

The visibility index is the cornerstone of our algorithm and we would like to clarify its behavior and the intuition behind it. Given a reference image $I_1$ and a distorted version of $I_1$ denoted with $I_2$, MSSIM index measures the similarity between $I_1$ and $I_2$.  The maximum value of the MSSIM index is $1$ and is obtained when $I_1$ and $I_2$ are identical, meaning that $I_2$ is not distorted at all. In our case, $I_s^{l, DAF}$ is the reference image and $I^{DAF}$ is considered a distorted version of $I_s^{l, DAF}$ because of the presence of the user's keypad. The maximum value of the visibility index is $1$ and is obtained when the user's keypad is completely out of perception. In figure \ref{fig:fig_22}, we demonstrate the behavior of the visibility index for an example hybrid keypad $I$. In the first row, we depict the third button of $I_s^{l, DAF}$, when $I_s^l$ is directly viewed from different distances. In the second row, we depict the third button of $I^{DAF}$, when $I$ is directly viewed from the same distances as $I_s^l$. In the third row, we provide the value of the visibility index for each viewing distance. From left to right, the viewing distance is increasing. As we can see, as the viewing distance is increasing, the digit $9$ which belongs to the user's keypad is fading away and the visibility index is increasing. When the visibility index becomes big enough, the digit from the user's keypad is no longer visible. We would like to make clear that we apply the MSSIM index between separated buttons and not between $I_s^{l, DAF}$ and $I^{DAF}$ as a whole, because in $I_s^{l, DAF}$ and $I^{DAF}$ exist big, almost identical, regions and the MSSIM index would have a very big value irrespectively of the buttons form.

The MSSIM index follows the premise that the main function of the human eye is to extract structural information from the viewing field. This connection to human perception is the main reason that we decided to use the MSSIM index. An additional advantage is that MSSIM index is very easily computed.

\subsubsection{Threshold Value of the Visibility Index}
\label{subsubsec:threshold}
Let's assume that we are given a hybrid keypad $I$ and an observer who first views $I$ from position $N_1$ and then from position $N_2$. If the corresponding visibility index values are $v_1$ and $v_2$ and holds $v_2 > v_1$, we expect the user's keypad to be less visible from position $N_2$ than from $N_1$. If $v_1 \simeq v_2$, we expect the user's keypad to be almost equally visible in both cases. This is a direct consequence of the way we have defined the visibility index. Now let's assume that two different hybrid keypads $I_1$ and $I_2$ are viewed by the same observer from positions $N_1$ and $N_2$, respectively. If the corresponding visibility index values are $v_1$ and $v_2$ and holds $v_2 > v_1$, we expect the user's keypads of $I_1$ to be more clearly visible than that of $I_2$. Similarly, if $v_1 \simeq v_2$, we expect the user's keypads of $I_1$ and $I_2$ to be almost equally visible. This is the main assumption that we make about the behavior of the visibility index and we expect to hold in its reverse form too. This means that if the user's keypad from a hybrid keypad $I_1$ is more clearly visible than the user's keypad of a different hybrid keypad $I_2$ when they are viewed from positions $N_1$ and $N_2$, respectively, then for the corresponding visibility index values $v_1$ and $v_2$ we expect to hold $v_2 > v_1$. If the user's keypads from $I_1$ and $I_2$ are almost equally visible, then we expect $v_1 \simeq v_2$.
It is important to mention that we expect these assumptions to hold only for the same observer. The reason is that the visual capabilities of different observers vary. For example, if a person with strong vision is directly viewing a hybrid keypad from a particular distance and is able to recognize the user's keypad with difficulty, then a person with weaker vision will have to go closer to the image to interpret it in the same way. As a result, the hybrid keypad will be interpreted in the same way by the two observers, but the value of the visibility index will be different.

Based on the aforementioned remarks, we set as a threshold $v_{th}$ the value of the visibility index when a particular observer is able to marginally recognize the digits of a user's keypad. Then, the visibility algorithm calculates the visibility index $v$ for the inputs $I$ and $N$, and compares it with $v_{th}$. If $v \geq v_{th}$, we predict that the user's keypad cannot be interpreted by the observer. If $v < v_{th}$, we predict that the observer is able to interpret the digits of the user's keypad. Since the threshold value will vary for different observers, we universally use the $v_{th}$ value that corresponds to people with the strongest vision, because we don't want to mistakenly predict that the user's keypad is not visible when it is.

\subsection{Parameter Tuning}
The parameters of the visibility algorithm are the spatial frequency $|\vec{f_1}|$ and the threshold $v_{th}$. To specify their values, we conducted a user study.

\subsubsection{Data Collection}
\label{subsubsec:data_collection}
\textbf{Participants.} We recruited $11$ participants from our institution, who were between $21$ and $34$ years of age. Our aim was to have participants with strong vision and that's why they all were of young age. Out of the $11$ participants, $6$ reported that they had either myopia or astigmatism, but they were wearing their glasses during the process. The rest of the participants reported that they did not have any problem.

\textbf{Materials.} We used $2$ different phones, a $Nexus \; 6$ and an $iPhone \; 6$. These two phones have displays with the same dimension ratio ($= \sfrac{16}{9}$), but different size and different resolution. The $Nexus \; 6$ has display size $5.96$ inches and resolution $1440 \times 2560$, while the $iPhone \; 6$ has display size $4.7$ inches and resolution $750 \times 1334$ pixels. For each phone, we created $7$ different categories of hybrid keypads. In $6$ of them, we set $\sigma_y^{lf} = 35 \; \sfrac{c}{im}$, since this is the value that we have decided to use in our hybrid keypads. The value of $\sigma_y^{hf}$ was equal to $120$, $160$, $200$, $240$, $280$ and $320 \; \sfrac{c}{im}$. In the seventh category, we tried a different value for $\sigma_y^{lf}$, which we set equal to $60 \; \sfrac{c}{im}$, while $\sigma_y^{hf} = 200 \; \sfrac{c}{im}$. For each category, we created $30$ hybrid keypads which had user's keypads with different digit ordering.

\textbf{Procedure.} Each subject participated in at least $3$ sessions. Each session was split in $3$ trials. The goal of each trial was to specify a viewing position $N$, from which the user's keypad of a hybrid keypad was marginally recognized. The hybrid keypad was displayed on a smartphone device. To specify the viewing position $N$ of the participant, we used spherical coordinates; meaning that $N = (r_0, \theta_0, \phi_0)$. The coordinate system was placed at the center of the image. In each trial, we kept $\theta_0$ and $\phi_0$ constant and we varied $r_0$.

The exact procedure was the following. During each trial, a smartphone device was placed on a tripod that matched the height of the participant. This way, when the participant was looking at the phone, $\theta_0$ was $\sfrac{\pi}{2}$ rad. To change the value of $\theta_0$, we tilted the phone on the tripod. Also, the participant was able to move relatively to the tripod in order to change the value of $\phi_0$. Having specified $\theta_0$ and $\phi_0$, the participant started to approach the phone from a big distance $r_0$. The initial value of $r_0$ was big enough to keep the user's keypad out of perception. As the participant was approaching to the phone, the user's keypad was starting to become visible. We recorded the maximum $r_0$ from which the participant was able to read the digits on the user's keypad. In particular, the participant had $15$ seconds to read all the $10$ digits and only one mistake was allowed. The time limit is connected to the fact that a shoulder-surfer has limited time to observe the user while entering her authentication credentials. Also, as the participant was approaching  the phone, each time he or she failed to read the digits, we switched to a different hybrid keypad from the same category in order to be sure that the participant is not getting familiar with a specific digit ordering. This is the reason that we created multiple hybrid keypads from each category for each phone.

For each session, the value of $\theta_0$ was constant. Specifically, $\theta_0$ was equal to $\sfrac{\pi}{2}$, $\sfrac{\pi}{4}$ or $\sfrac{\pi}{3} \; rad$. On each trial of a session, we had a different value for $\phi_0$. During the first trial, we had $\phi_0 = 0 \; rad$, during the second $\phi_0 = \sfrac{\pi}{6} \; rad$, and during the third $\phi_0 = \sfrac{\pi}{3} \; rad$. Since $\theta_0 \in (0, \pi)$ and $\phi_0 \in (-\sfrac{\pi}{2}, +\sfrac{\pi}{2})$, the values of $\theta_0$ and $\phi_0$ that we considered belonged to the $\sfrac{1}{4}$ of the available space. The reason is that the visual angle is symmetric with respect to the $xy$, $zx$ and $zy$ planes as can be easily seen in equations \ref{eq:7} and \ref{eq:8}. In addition, on each session we used hybrid keypads from a single category and we used only one smartphone device.

Last but not least, we have to comment on the illumination of the scene since it affects the perception of the hybrid keypad. The experiments were conducted in an indoor space with normal artificial lighting and the smartphone was carefully placed to avoid distracting reflections from surrounding objects. In addition, the brightness of the screen was set to its maximum value. In general, we tried to create favorable conditions for the observers, since the collected data would be used for the estimation of the security strength of IPIN, and we wanted to correspond to the worst case scenario.

\textbf{Collected data.} An entry in our dataset is composed of the identification number of a participant, the hybrid keypad that was at view, and the viewing position from which the participant was able to marginally recognize the digits from the user's keypad. In total, we have $270$ entries. We tried to balance the number of data that we collected with each phone and from each hybrid keypad category. We stopped collecting data when the statistical analysis provided robust results with respect to the variables of our data.

\subsubsection{Estimation of the DAF Filter Parameters}
\label{subsubsec:DAF}
\begin{figure}[!t]
\centering
\includegraphics[width=3.5in]{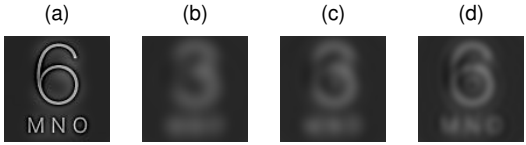}
\caption{(a) The third button from a hybrid keypad. We simulate the perception of this button from position $(63, \sfrac{\pi}{2}, 0)$ by using DAF filters with different $|\vec{f_1}|$ values. (b) The filtered button with $|\vec{f_1}| = 25 \; \sfrac{c}{d}$. (c) The filtered button with $|\vec{f_1}| = 31 \; \sfrac{c}{d}$. (d) The filtered button with $|\vec{f_1}| = 42 \; \sfrac{c}{d}$.}
\label{fig:fig_24}
\end{figure}
We considered a subset of the data with $81$ entries from $3$ participants and we calculated the visibility index for each entry. We expected the behavior of the DAF filter to have two characteristics. The first was the visibility index to have small standard deviation for each participant. The second was the filtered hybrid keypads to have a user's keypad with digits that can be marginally recognized. Loftus et al. \cite{loftus2005easier} suggested $3$ possible values for $|\vec{f_1}|$, depending on the task at hand. Unfortunately, there was not a unique value for $|\vec{f_1}|$ that could create a DAF filter with the desired behavior for our task. The reason was that when the viewing position $N = (r_0, \theta_0, \phi_0)$ of a participant had big $\phi_0$ or small $\theta_0$, the corresponding value of the visibility index was significantly smaller. This means that when the participants were viewing a hybrid keypad from an angle, they had to approach closer to the screen than expected according to the visibility index values. We assume that this problem is connected with the fact that the brightness of LCD screens, like the ones used by $iPhone \; 6$ and $Nexus \; 6$, dims when an observer is viewing from an angle. As a result, when the participants were viewing from an angle, the perceived contrast was lower and more gratings from the hybrid keypad were not perceived. To account for this effect, we gradually reduce the value of $|\vec{f_1}|$ as $\phi_0$ is increasing and $\theta_0$ is decreasing. Based on these remarks, we defined the DAF filter in $2$ steps.

In the first step, the goal was to specify the initial value of $|\vec{f_1}|$ when the brightness of the screen is unaffected by the viewing angle. The possible values for $|\vec{f_1}|$ were $25$, $31$ and $42 \; \sfrac{c}{d}$, as proposed by Loftus et al. \cite{loftus2005easier}. To select the most suitable value, out of the $81$ entries in our dataset, we singled out those with $\phi_0 = 0$ and $\theta = \sfrac{\pi}{2}$, meaning the entries that corresponded to an observer who is looking directly at the screen. Then, since this subset had a small size, we applied the $3$ DAF filter variants to the corresponding hybrid keypads and we picked the filter that created hybrid keypads with marginally recognizable user's keypad. Figure \ref{fig:fig_24} (a) provides the third button of a hybrid keypad from our subset. The digit of the shoulder-surfer's keypad is $3$ and the digit of the user's keypad is $6$. We simulate how the button is perceived when it is directly viewed from $63$ inches, since this viewing position was part of our data for this hybrid keypad. In figure \ref{fig:fig_24} (b), we used a DAF filter with $|\vec{f_1}| = 25 \; \sfrac{c}{d}$, in figure \ref{fig:fig_24} (c), $|\vec{f_1}| = 31 \; \sfrac{c}{d}$, and in figure \ref{fig:fig_24} (d), $|\vec{f_1}| = 42 \; \sfrac{c}{d}$. As we can see, the digit on the button is interpreted as $3$ in figure \ref{fig:fig_24} (b), it is marginally interpreted as $6$ (or even $3$) in figure \ref{fig:fig_24} (c), and is clearly interpreted as being $6$ in figure \ref{fig:fig_24} (d). Following this way of reasoning, we concluded that the most suitable initial value for $|\vec{f_1}|$ is $31 \; \sfrac{c}{d}$.

In the second step, we modeled how the value of $|\vec{f_1}|$ changes as a function of $\theta_0$ and $\phi_0$. We assume that $|\vec{f_1}| = 31 \cdot A(\phi_0, \theta_0)$. For the function $A$, we have:
\begin{equation} \label{eq:15}
A(\phi_0, \theta_0) = \Bigg(1-\Big(\frac{\phi_0}{\sfrac{\pi}{2}}\Big)^{k_a}\Bigg) \cdot \Bigg(1-\Big(\frac{\theta_0 - \sfrac{\pi}{2}}{\sfrac{\pi}{2}}\Big)^{k_a}\Bigg)
\end{equation}
where $k_{a}$ is a positive real number that controls how fast the value of $|\vec{f_1}|$ drops when $\phi_0$ is increasing or $\theta_0$ is decreasing. We use the same number $k_a$ in both factors because we assume that the brightness of the screen changes in the same way when either $\phi_0$ or $\theta_0$ is changing. We considered five different cases for the effect of $\phi_0$ and $\theta_0$. The first was that they don't affect $|\vec{f_1}|$, meaning that $A = 1$. The rest four cases corresponded to $k_{a} = 2.5$, $k_{a} = 3$, $k_{a} = 3.5$ and $k_{a} = 4$. For each of these cases, we created the corresponding DAF filter and we computed the visibility index for all the $81$ entries in our subset. Then, for each participant we calculated the standard deviation of the visibility index. In Table \ref{tab:tab_1} we provide the results from this process. Each row corresponds to a different participant and each column corresponds to a different function $A(\phi_0, \theta_0)$. As we can see, all the participants demonstrated the smallest standard deviation for $k_{a} = 3$, and consequently, this was the choice we made. We have to note that the data we used were from participants without reported problems in their vision, because we wanted to increase the probability of defining a DAF filter that simulates the visual perception of a person with strong vision.
\begin{table}
\caption{The standard deviation of the visibility index for three participants $p_1$, $p_2$ and $p_3$, when different forms of the function $A(\phi_0, \theta_0)$ are used. The minimum standard deviation value for each participant is highlighted.}
\label{tab:tab_1}
\centering
\begin{adjustbox}{max width=\textwidth}
\begin{tabular}{|c|c|c|c|c|c|} 
\hline
& $A = 1$ & $k_{a} = 2.5$ & $k_{a} = 3$ & $k_{a} = 3.5$ & $k_{a} = 4$ \\ 
\hline
$p_1$ & 0.0496 & 0.0181 & \textbf{0.0138} & 0.0159 & 0.0212 \\ 
\hline
$p_2$ & 0.0546 & 0.0167 & \textbf{0.0112} & 0.0141 & 0.0223 \\
\hline
$p_3$ & 0.0560 & 0.0224 & \textbf{0.0178} & 0.0183 & 0.0240 \\
\hline
\end{tabular}
\end{adjustbox}
\end{table}

\subsubsection{Estimation of the Threshold Value for the Visibility Index}
\label{subsubsec:estimation_threshold}
\begin{table*}
\caption{The coefficient of variation for the visibility index values of each participant in our dataset.}
\label{tab:tab_3}
\centering
\begin{adjustbox}{max width=\textwidth}
\begin{tabular}{|c|c|c|c|c|c|c|c|c|c|c|c|} 
\hline
& $p_1$ & $p_2$ & $p_3$ & $p_4$ & $p_5$ & $p_6$ & $p_7$ & $p_8$ & $p_9$ & $p_{10}$ & $p_{11}$ \\
\hline
$cv \; (\%)$ & 1.85 & 1.51 & 1.48 & 1.92 & 2.01 & 5.41 & 2.66 & 2.06 & 2.54 & 2.51 & 3.86 \\
\hline
\end{tabular}
\end{adjustbox}
\end{table*}
According to the assumptions we made in Section~\ref{subsubsec:threshold}, the visibility index should have almost identical values for each participant in our dataset, irrespectively of the phone that was used during the data collection process, or the categories that the hybrid keypads belonged to. To verify this, we used the DAF filter we defined in the previous section to compute the visibility index for each of the $270$ entries in our dataset. Then, we computed the coefficient of variation (cv) for the visibility index values of each participant. We provide the results in Table \ref{tab:tab_3}. As we can see, the cv is lower than $10 \%$ for every participant, and consequently, we concluded that the values of the visibility index are homogeneous for each participant.

We wanted to further test the assumption that the values of the visibility index for each participant are independent of the smartphone device that was used. Also, we wanted to test the assumption that the values of the visibility index vary significantly between different participants. We tested both of these assumptions by applying a two-factor ANOVA with randomized complete block design. We used data from the participants who were exposed to both phones during the data collection process. These were $5$ out of the $11$ participants. The visibility index values from different participants were assigned to separate blocks, and consequently, the participants were the blocking factor. The factor of interest within each block was the type of the phone; $Nexus \; 6$ or $iPhone \; 6$. Since the normality and homoscedasticity conditions were satisfied, we were able to compute the p-value of the two factors. The p-value for the type of the phone was $12.07 \% > 5 \%$ and we concluded that the smartphone device is not a statistically significant factor of variation for the value of the visibility index. In contrast, the p-value for the participants was less than $1 \permil$ and consequently, the values of the visibility index between the participants have statistically significant differences.
\begin{figure}[!t]
\centering
\includegraphics[width=2.5in]{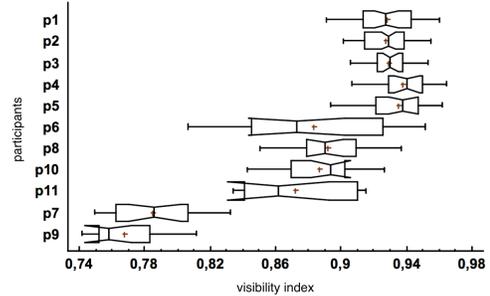}
\caption{The box and whisker plot with median notch after applying the Kruskal-Wallis test in our whole dataset with the participants as the factor of variation.}
\label{fig:fig_28}
\end{figure}

We further tested if participants are a factor of variation by considering the whole dataset with the $11$ participants. Our intention was to apply an one-way ANOVA, but the normality and homoscedasticity conditions weren't satisfied, and as a result, we applied the non-parametric Kruskal-Wallis test. The p-value was less than $1 \permil$ and we confirmed that there are statistically significant differences between the visibility index values of different participants. In figure \ref{fig:fig_28}, we provide the corresponding box and whisker plot with median notch of the visibility indexes of all participants. As we can see, the participants formed three main groups based on their visibility index values. Participants $p1$, $p2$, $p3$, $p4$ and $p5$ formed the first group, $p6$, $p8$, $p10$ and $p11$ formed the second group, and $p7$ and $p9$ formed the third group. We confirmed this grouping of the participants by performing pairwise comparisons with Man-Whitney tests. Each group had participants with visibility index values in a different range and this is because the participants demonstrated different visual capabilities. The first group had the participants with the highest visibility index values and consequently, these were the people who demonstrated the strongest vision. The mean value of the visibility index in this group was $0,92996$ with standard error $0.00115$. The desired threshold value of the visibility index was set equal to $0.93$.

\section{Safety Distance}
\label{sec:safety_distance}
\subsection{Shoulder-Surfing With a Naked Eye}
\begin{figure}[!t]
\centering
\includegraphics[width=3.5in]{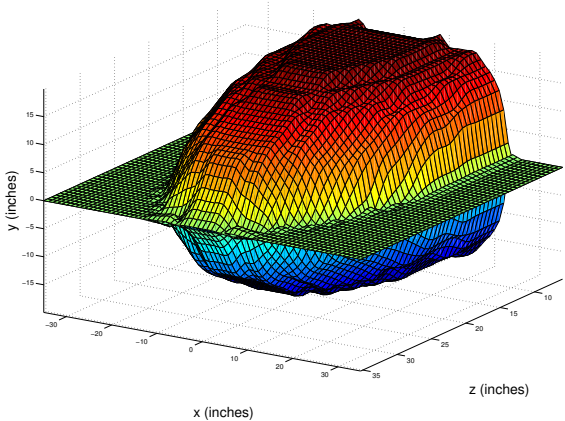}
\caption{The visibility region of a hybrid keypad $I$ created with $\sigma_y^{lf} = 35 \; \sfrac{c}{im}$ and $\sigma_y^{hf} = 320 \; \sfrac{c}{im}$ for the $iphone \; 6$ smartphone.}
\label{fig:fig_30}
\end{figure}
We assume that we are given a hybrid keypad $I$ and an observer at a position $N = (r_0, \theta_0, \phi_0)$. If $d_s$ is a safety distance for $I$, then $\forall \theta_0,
 \phi_0$, if $r_0 \geq d_s$, the user's keypad can not be interpreted by the observer. Of course, we are interested in the minimum possible value of $d_s$, because it corresponds to the maximum protection that $I$ can offer against shoulder surfing. In addition, as we have seen in previous sections, even in the case that we are given a desired $d_s$ and we are asked to design a hybrid keypad $I$ that satisfies it, we aim to maximize the usability of $I$ by making $d_s$ to be the minimum safety distance of $I$. So, we are only interested in the minimum possible value of the safety distance.

To estimate the minimum $d_s$ for a hybrid keypad $I$, we first examine from which viewing positions at the 3D space the user's keypad of $I$ is visible. We call the resulting region of the 3D space, ``visibility region". To estimate the visibility region of a hybrid keypad, we applied the visibility algorithm for viewing positions from a dense grid at the 3D space. For a hybrid keypad $I$ created with $\sigma_y^{lf} = 35 \; \sfrac{c}{im}$ and $\sigma_y^{hf} = 320 \; \sfrac{c}{im}$ for the $iphone \; 6$ smartphone, the visibility region is depicted in Figure \ref{fig:fig_30}. The depicted $zx$ plane can be used as a reference to understand the geometry of the visibility region. The user's keypad is visible when $I$ is viewed from positions which are either on the depicted surface or they are enclosed by it. The coordinate system is the same as the one used in Figure \ref{fig:fig_5}, with the image placed in position $(0, 0, 0)$. In addition, for all viewing positions $N = (x_0, y_0, z_0)$, we assume that $z_0 > 5$ inches. The visibility region is symmetric with respect to planes $zx$ and $yz$. This is something that we expected because of the symmetry in equations \ref{eq:7} and \ref{eq:8}. In general, the form of the visibility region is similar to that of an ellipsoid. If we consider spherical coordinates, the safety distance will be slightly bigger than the biggest viewing distance $r_0$ of a position $N = (r_0, \theta_0, \phi_0)$ that belongs to the visibility region. Since the form of the visibility region is ellipsoidal, for the position $N$ with the maximum $r_0$ will hold $\theta_0 = \sfrac{\pi}{2}$ and $\phi_0 = 0$. This is a result that we expected, since for a given $r_0$, from equations \ref{eq:12} and \ref{eq:13}, we can easily derive that the visual angle is maximized when $\theta_0 = \sfrac{\pi}{2}$ and $\phi_0 = 0$.

In our threat model, we have assumed that for the viewing position $N = (r_0, \theta_0, \phi_0)$ of the shoulder-surfer holds $|\phi_0| \geq \sfrac{\pi}{6} \; rad$. In the visibility region, as $|\phi_0|$ is increasing, the maximum $r_0$ is decreasing. As a consequence, we define the safety distance to be the minimum distance $d_s$ for which holds that from the viewing position $N = (r_s, \sfrac{\pi}{2}, \sfrac{\pi}{6})$, an observer with strong vision is unable recognize the digits on the user's keypad. Based on this definition, if we are a given a hybrid keypad $I$, we can apply the visibility algorithm for viewing positions $N = (r_0, \sfrac{\pi}{2}, \sfrac{\pi}{6})$ with varying $r_0$, and set the safety distance to be equal to the minimum $r_0$ for which holds that the corresponding visibility index $v$ is greater than $v_{th}$. We are also interested in the case that we are given as a security requirement a safety distance $d_s$ and we have to set the value of $\sigma_y^{hf}$. To do this, we vary the value of $\sigma_y^{hf}$ that we use to create $I$, and we apply the visibility algorithm for the position $N = (d_s, \sfrac{\pi}{2}, \sfrac{\pi}{6})$. We create $I$ with the minimum $\sigma_y^{hf}$ for which holds that the visibility index $v$ is greater than $v_{th}$. This way we were able to create the $4$ hybrid keypads that we provide as predefined options to the users of IllusionPIN.


\subsection{Shoulder-Surfing Through a Surveillance Camera}
We assume that we are given a hybrid keypad $I$, a smartphone device where $I$ is displayed on, and a surveillance camera at position $N = (r_0, \theta_0, \phi_0)$. To estimate the safety distance, we calculate the minimum $r_0$ for which holds that $\forall \theta_0, \phi_0$, the camera is unable to capture the user's keypad $I_u^h$. We assume that the camera is unable to capture $I_u^h$ when a cycle from the grating of $I_u^h$ with the biggest cycle size, occupies at most a pixel when it is projected on the image plane of the camera.

We start by estimating the smallest spatial frequency components which are present in $I_u^h$, since they will correspond to the cycles with the biggest size. We remind that $I_u^h$ is the result of applying the high-pass filter $G_h$ to $I_u$. We assume that when $G_h$ assigns a gain value less than $0.5$ to a spatial frequency, the corresponding grating is cut off. The isocontour of $G_h$ that corresponds to the value $0.5$ will be an ellipse that we call the cut-off ellipse.
We consider that all the spatial frequencies enclosed by the cut-off ellipse are cut off. Since $G_h = 1 - G^h_l$, where $G^h_l(f_x, f_y) = exp(-\sfrac{1}{2 \sigma_x^{hf}} \cdot f_x^2 -\sfrac{1}{2 \sigma_y^{hf}} \cdot f_y^2)$, for the axes of the cut-off ellipse will hold:
\begin{equation} \label{eq:16}
a = \sqrt{2 \sigma_x^{hf} \log{(\sfrac{1}{0.5})}}
\end{equation}
\begin{equation} \label{eq:17}
b = \sqrt{2 \sigma_y^{hf} \log{(\sfrac{1}{0.5})}}
\end{equation}

\noindent
To describe the region of the spatial frequencies which are cut off by the filter in a simpler way, we consider the rectangle with the biggest area that is inscribed to the cut-off ellipse. We assume that the spatial frequencies which are cut off by the filter, are those enclosed by this rectangle instead of those enclosed by the cut-off ellipse. We call this rectangle, the cut-off rectangle. For the biggest horizontal and vertical spatial frequency components in the cut-off rectangle will hold:
\begin{equation} \label{eq:18}
f^s_x = a \cdot \sfrac{\sqrt{2}}{2}
\end{equation}
\begin{equation} \label{eq:19}
f^s_y = b \cdot \sfrac{\sqrt{2}}{2}
\end{equation}

\noindent
Based on these remarks, we conclude that for a grating from $I_u^h$ with spatial frequency $\vec{f} = (f_x, f_y)$ will hold $f_x \geq f^s_x$ and $f_y \geq f^s_y$. For an example filter $G_h$ with $(\sigma_x^{hf}, \sigma_y^{hf}) = (120.9375 \; \sfrac{c}{im}, 215 \; \sfrac{c}{im})$, we calculate $f^s_x = 100.69 \; \sfrac{c}{im}$ and $f^s_y = 178.99 \; \sfrac{c}{im}$.

After measuring the smallest frequency components present in $I_u^h$, we should calculate the length of the corresponding cycles. To this end, we need to know the resolution $x_r \times y_r$ of the screen that is used to display $I$ and the number of pixels per inch ($ppi$). Based on that, we can calculate the biggest cycle size in each dimension:
\begin{equation} \label{eq:20}
l_x = \frac{\sfrac{x_r}{ppi}}{f^s_x}
\end{equation}
\begin{equation} \label{eq:21}
l_y = \frac{\sfrac{y_r}{ppi}}{f^s_y}
\end{equation}

\noindent
where $\sfrac{x_r}{ppi}$ and $\sfrac{y_r}{ppi}$ express the length of the display in the horizontal and vertical dimension respectively, measured in inches. If we assume that $I$ is displayed on a $Nexus \; 6$ device, the resolution is $1440 \times 2560 \; pixels$ and $ppi = 493 \; \sfrac{p}{in}$. As a consequence, we find that $l_x = 0.029 \; inches$ and $l_y = 0.029 \; inches$. As we can see, both lengths are equal. This is something that we expected, because we have designed $G_h$ to be isotropic in the spatial domain.

To estimate the safety distance $d_s$, we assume that we have a square stimulus $c$ of size $(l_x, l_y)$, which is viewed by a camera in the same way that an image is viewed by a human observer in Figure \ref{fig:fig_5}. This means that the camera is modeled as an ideal pinhole camera. In this setting, the camera is in position $N = (r_0, \theta_0, \phi_0)$. As we increase $r_0$, $c$ subtends increasingly smaller angle.
The camera will not be able to capture $c$ for the first time when $c$ will subtend visual angle that corresponds exactly to one pixel. As a result, safety distance will be the biggest distance for which holds that $c$ is projected to exactly one pixel. From equations \ref{eq:12} and \ref{eq:13}, we can easily see that a specific visual angle corresponds to the biggest $r_0$ when $\theta_0 = \sfrac{\pi}{2}$ and $\phi_0 = 0$. As a result, in our setting, we assume that the camera is in position $N = (d_s, \sfrac{\pi}{2}, 0)$, while $c$ subtends visual angle that corresponds to exactly one pixel on the image plane.
Since $c$ is a square stimulus,
we simply use perspective projection in the horizontal dimension to find that for $d_s$ holds:
\begin{equation} \label{eq:22}
d_s = f \cdot \sfrac{l_x}{s_p}
\end{equation}

\noindent
where $f$ is the focal length of the camera, and $s_p$ is the pixel size. If we assume that the surveillance camera has the specifications of a $Nexus \; 6$ camera, for the pixel size will hold $s_p = 0.001127 \; mm$ and for the focal length $f = 3.8 \; mm$. As a result, we get $d_s = 97.81 \; inches$.

In our threat model, we assumed that for this attack scenario the surveillance camera is recording from a distance that is at least $100 \; inches$. According to the previous calculations, for a hybrid keypad created with $(\sigma_x^{hf}, \sigma_y^{hf}) = (120.9375 \; \sfrac{c}{im}, 215 \; \sfrac{c}{im})$ and displayed on a $Nexus \; 6$ smartphone, our security requirement is satisfied. It is important to note that this specific keypad is the second most usable hybrid keypad out of the $4$ predefined options that we offer to $Nexus \; 6$ users. In addition, we have to consider that all the assumptions that we made during our calculations were in favor of the attacker. Some examples are that we required the camera to be unable to capture all the gratings from $I_u^h$ and not a subset of them, that we disregarded the effect of $I_s^l$, and that we overestimated the specifications of the surveillance camera. All these show that under real life conditions, is extremely improbable to capture the user's keypad of IllusionPIN with a surveillance camera.

\section{Evaluation}
\label{sec:evaluation}
We would like to estimate the probability of underestimating the safety distance with the visibility algorithm. In other words, we would like to know how probable it is for a shoulder-surfer to steal the credentials of an IllusionPIN user, even if the viewing distance of the attacker is equal or bigger than the estimated safety distance. To this end, we performed simulated shoulder-surfing attacks against IllusionPIN.

\textbf{Participants.} We recruited $21$ participants who were undergraduate and graduate students from our institution. All participants were less than $40$ years old and they had either normal or corrected vision. Note that all experiments in this work were approved by the IRB of our institution.

\textbf{Materials.} We built an application for the Android operating system to aid the execution of the simulated attacks. The application was run on a $Nexus \; 6$ phone and allowed the users to create five keypad categories that we denote with $c_i$, for $i = 1,2,3,4,5$. The first $4$ categories were hybrid keypads that corresponded to the $4$ safety distance values $d_s^i$, $i = 1,2,3,4$, that we have considered in our threat model. All hybrid keypads were created with $\sigma_y^{lf} = 35 \; \sfrac{c}{im}$. Category $c_1$ was created with $\sigma_y^{hf} = 145  \; \sfrac{c}{im}$ and corresponded to $d_s^1 = 60$ inches, $c_2$ had $\sigma_y^{hf} = 215 \; \sfrac{c}{im}$ and $d_s^2 = 45$ inches, $c_3$ had $\sigma_y^{hf} = 305 \; \sfrac{c}{im}$ and $d_s^3 = 35$ inches, and $c_4$ had $\sigma_y^{hf} = 440 \; \sfrac{c}{im}$ and $d_s^4 = 25$ inches. Category $c_5$ was a normal keypad that is used in regular PIN authentication.

\textbf{Procedure.} Participants worked in pairs. Each pair performed $10$ simulations in total. In the first $5$ simulations one participant was playing the role of the attacker and the other the role  of the user. Then, the participants were asked to switch roles and repeat the same $5$ simulations. The first $5$ simulations $s_i$, $i = 1,2,3,4,5$, were performed in the following way. The shoulder-surfer was placed at position $N = (d_s^i, \sfrac{\pi}{2}, \sfrac{\pi}{6})$, while the phone was at position $N = (0, \sfrac{\pi}{2}, 0)$. The user freely created a $4$-digit PIN with a keypad from category $c_i$, and was asked to successfully authenticate $3$ times. The attacker was able to observe the user during all authentication attempts and then was asked to replicate the PIN. During $s_5$, we set $d_s^5 = d_s^1$, which was the biggest distance that we considered.

Before the simulations, we asked the participants to practice with all the keypad categories by authenticating $10$ times with each category. During the simulations, we tried to create favorable conditions for the attackers. In particular, the simulations were performed in a silent indoor place with adequate illumination, while the brightness of the phone screen was set to its maximum level. In addition, the users were asked to keep the screen of the phone in the view of the attackers during the authentication. It is also important to clarify that we purposely selected participants of young age with normal or corrected vision, because these participants could impersonate skillful shoulder-surfers.

\textbf{Collected data.} We simulated $84$ shoulder-surfing attacks against IllusionPIN, and none of them was successful. The best performance was demonstrated by $2$ participants who managed to correctly replicate $3$ out of $4$ PIN digits in one attack each. All $21$ attacks against the regular PIN authentication were successful.

\textbf{Data analysis.}
It is very important that none of the attackers was able to break IllusionPIN.
We should also note that the category of a hybrid keypad wasn't a factor that affected the success rate of the attackers. Based on that, we consider that we have $21$ independent samples and we calculate the Clopper-Pearson interval for the success rate of shoulder-surfing attacks. With probability $95 \%$, we expect the interval $[0, 0.1329]$ to contain the success rate. If we increase the sample size, the range of the interval will be reduced. However, based on the unfavorable conditions for our scheme, the visual capabilities of the participants, and the performance of the attackers, we expect that we have to record a very large number of attacks to find a successful one. As a result, we are already confident that the success rate is much closer to $0$ than to $0.1329$.

\section{Limitations and Concluding Remarks}
\label{sec:limitation}
The main goal of our work was to design a PIN-based authentication scheme that would be resistant against shoulder-surfing attacks. To this end, we created IllusionPIN. We quantified the level of resistance against shoulder-surfing by introducing the notion of safety distance, which we estimated with a visibility algorithm. In the context of the visibility algorithm, we had to model at a basic level how the human visual system works. In this process, we made a number of simplifying assumptions that limit the accuracy of our calculations. The most obvious example is the pinhole camera model that we used to describe the image formation process in the eye. This is a widely used model, but disregards important parts of the human eye, like the lens. In Section~\ref{subsubsec:distance}, we made an additional simplification by assuming that the perception of an image depends on the visual angle that it subtends, no matter where the observer stands. A problem with this assumptions is that, depending on the viewing position, we may perceive the dimensions of an image as having a different ratio. For example, as we can see in equations \ref{eq:12} and \ref{eq:13}, when $\phi_0$ is increasing, only $\theta_x$ is decreasing and the image is perceived as being squeezed in the horizontal dimension. As a consequence, the image of a circle could be perceived as the image of an ellipse when it is viewed from a big angle. It is interesting to note that even if we did not explicitly model such  phenomena, the factor $A(\phi_0, \theta_0)$ that we estimated in Section~\ref{subsubsec:DAF} may have accounted for them implicitly to some extent. It is very important that despite all these simplifying assumptions that we made, our results led us to conclusions that agreed with our expectations. For example, the visibility region depicted in Figure \ref{fig:fig_30} has the expected ellipsoidal form, while the visibility index demonstrated the behavior we described in Section~\ref{subsubsec:threshold}. So, more strict assumptions, followed by more detailed models, could improve the accuracy of our current results, but we expect the general conclusions to remain the same.
\begin{figure}[!t]
\centering
\includegraphics[width=0.7in]{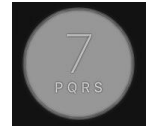}
\caption{A button from a different kind of hybrid keypad. The shoulder-surfer's keypad is composed of uniformly white buttons.}
\label{fig:fig_33}
\end{figure}

The visibility algorithm forms the core of our work and we would like to examine whether it can be used to assess the visibility of images other than hybrid keypads. The visibility algorithm uses the MSSIM index which quantifies the distortion between two images. If we are given a random image $I_r$ and a viewing position $N$, we could apply the same rationale by considering $I_r^{DAF}$ as the distorted version of $I_r$. 
However, we do not expect the visibility index threshold value that we specified in Section~\ref{subsubsec:estimation_threshold} to be applicable to random images. The reason is that the level of distortion does not uniquely correspond to how visible particular visual details are in the distorted image. To verify this, we extended the data collection process that we presented in Section~\ref{subsubsec:data_collection} to a different kind of hybrid keypads that we call white keypads. An example button from such a keypad is depicted in Figure \ref{fig:fig_33}. As we can see, the shoulder-surfer's keypad of a white keypad is composed of all white buttons. In the small dataset that we collected, the visibility index values were consistent for a particular user, but they were considerably lower than the corresponding values that the same user reported in the original dataset. This means that even if a person perceives the digits on a hybrid keypad to be equally visible to the digits on a white keypad, the distortion in the white keypad is bigger and the visibility index has a lower value. This is something logical, because when the reference buttons are all white, a digit that is even slightly visible is considered a big distortion. Based on that, we conclude that the visibility index threshold value is not universal. We would also like to remind that the values of the DAF filter parameters depend on the task at hand. As a result, we may have to repeat the estimation process for a considerably different task. We conclude that the visibility algorithm could be used to assess the visibility of general images, but its parameters have to be appropriately tuned for the particular task at hand.
\section*{Acknowledgment}
The authors would like to thank all the participants of the conducted user studies for their cooperation and contribution. Also, they would like to thank Assoc. Prof. George Papadopoulos (Agricultural Univ. of Athens, Department of Crop Science, Laboratory of Biometry) for his contribution in the design and implementation of the statistical analysis presented in sections~\ref{subsubsec:estimation_threshold} and~\ref{sec:evaluation}.

%
\bibliographystyle{abbrv}

%
%
\balance

\end{document}